\documentclass[aps,prd,reprint,superscriptaddress,nofootinbib,amsmath,amssymb,floatfix,preprintnumbers]{revtex4-2}

\usepackage{graphicx,subfigure}
\usepackage{svg}
\svgsetup{
    inkscapepath=i/svg-inkscape/
}
\svgpath{{svg/}}
\usepackage{xcolor}
\usepackage{setspace}
\usepackage{hyperref}
\usepackage[utf8]{inputenc}
\usepackage[T1]{fontenc}
\usepackage{lmodern}

\usepackage{booktabs}
\usepackage{color}
\usepackage{epsfig}
\usepackage{ifpdf}
\usepackage{amsmath}
\usepackage{bm}
\usepackage[english]{babel}
\usepackage{amsfonts}
\usepackage{amssymb}
\usepackage{braket}
\usepackage{enumerate}
\usepackage{cancel}
\usepackage{multirow}
\usepackage{cleveref}
\usepackage{xspace}
\usepackage{array}
\usepackage{fancyvrb}
\usepackage{fontawesome}
\usepackage{dcolumn}
\usepackage{slashed}
\usepackage[normalem]{ulem}
\usepackage{url}

\usepackage[absolute,overlay]{textpos} 
\def\XXint#1#2#3{{\setbox0=\hbox{$#1{#2#3}{\int}$}
     \vcenter{\hbox{$#2#3$}}\kern-.5\wd0}}

\makeatletter
\g@addto@macro\bfseries{\boldmath}
\makeatother

\definecolor{nicered}{rgb}{0.7,0.1,0.1}
\definecolor{nicegreen}{rgb}{0.1,0.5,0.1}
\hypersetup{colorlinks, urlcolor=blue, citecolor=nicegreen,linkcolor= nicered}

\begin{document}



\title{Constraints on Axion-Like Particles with the Silicon Detector at a Nuclear Reactor}

\author{Yuanlin Gong}
\affiliation{Department of Physics and Institute of Theoretical Physics, Nanjing Normal University, Nanjing, 210023, China}

\author{Jun Guo}
\affiliation{College of Physics and Communication Electronics, Jiangxi Normal University, Nanchang 330022, China}

\author{Ning Liu}
\email{Corresponding author: liuning@njnu.edu.cn}
\affiliation{Department of Physics and Institute of Theoretical Physics, Nanjing Normal University, Nanjing, 210023, China}
\affiliation{Nanjing Key Laboratory of Particle Physics and Astrophysics, Nanjing, 210023, China}

\author{Liangliang Su}
\affiliation{Institute for Astroparticle Physics (IAP), Karlsruhe Institute of Technology (KIT),
Eggenstein-Leopoldshafen, 76344, Germany}

\author{Wen-Na Yang}
\email{Corresponding author: wennayang@njnu.edu.cn}
\affiliation{Department of Physics and Institute of Theoretical Physics, Nanjing Normal University, Nanjing, 210023, China}

\begin{abstract}

Axion and axion-like particles (ALPs), predicted in various extensions of the Standard Model, can be abundantly produced in nuclear reactors via the Primakoff process. In this work, we explore the detection of ALPs in silicon detectors through plasmon excitations. Owing to their relativistic nature, reactor-produced ALPs can efficiently excite plasmon resonances, while the accompanying energetic photon typically escapes from the thin detector without depositing an appreciable amount of energy. Utilizing the data from the Connie and Atucha-II experiments, we set the 90\% confidence level upper limits on the ALP-photon coupling $g_{a\gamma\gamma}$ over the axion mass range $0.1-100$ keV. We further show that, for an exposure of 30 kg$\cdot$yr, the projected sensitivity of vIOLETA exceeds the current NEON limit by a factor of three in the same mass range. This improvement would expand the explored region of the QCD axion and ALP parameter space.
\end{abstract}




\maketitle

\section{Introduction}
Axion and axion-like particles (ALPs) are a well-motivated class of light pseudoscalar bosons that emerge in a wide range of theories beyond the Standard Model  (SM) with broken global symmetry~\cite{Chikashige:1980ui,Wilczek:1982rv,Masso:1995tw,Branco:2011iw,Marsh:2015xka} and in string theory \cite{Arvanitaki:2009fg,Svrcek:2006yi}. They have gained significant attention both as portals to hidden sectors and as dark matter candidates~\cite{Preskill:1982cy,Abbott:1982af,Dine:1982ah,Duffy:2009ig,Boehm:2014hva,Berlin:2014tja,Dolan:2014ska,Arias:2012az}. The QCD axion provides perhaps the most elegant solution to the strong CP problem~\cite{Peccei:1977hh,Weinberg:1977ma,Wilczek:1977pj}, while the ALPs such as the relaxion may solve the hierarchy problem via the so-called relaxion mechanism \cite{Graham:2015cka,Flacke:2016szy} associated with electroweak symmetry breaking. ALPs give rise to a rich and uncharted phenomenological landscape, where both the ALP mass and its couplings to SM particles span many orders of magnitude~\cite{Svrcek:2006yi,Brivio:2017ije,DiLuzio:2020wdo,Arza:2025kuh,Guo:2025dlk}. 

Originally expected ALPs with mass $\mathcal{O}$(meV) have been most extensively investigated, as evidenced by abundant ongoing and proposed experiments \cite{Jaeckel:2010ni,Irastorza:2018dyq,Sikivie:2020zpn,Song:2024rru}. For example, ALP-photon couplings are explored by well-established experiments, including helioscopes (CAST~\cite{CAST:2017uph} and IAXO~\cite{IAXO:2019mpb}), haloscopes ( ABRACADABRA~\cite{Ouellet:2018beu}, ADMX ~\cite{ADMX:2019uok} and  HAYSTAC~\cite{HAYSTAC:2018rwy}), light-shining-through-walls experiments like ALPS II~\cite{Spector:2019ooq}, and interferometry~\cite{DeRocco:2018jwe,Obata:2018vvr,Liu:2018icu}.
Light ALPs with mass $\mathcal{O}$(keV) remain of significant interest due to their potential implications for Big Bang Nucleosynthesis~\cite{Masso:1997ru,Cadamuro:2010cz,Cadamuro:2011fd,Millea:2015qra,Depta:2020wmr}, anomalous astrophysical emission lines~\cite{Jaeckel:2014qea,Higaki:2014zua}, and unexplained nuclear resonances~\cite{Ellwanger:2016wfe,Krasznahorkay:2019lyl}. Although cosmological and astrophysical observations~\cite{Ayala:2014pea,Hardy:2016kme,Dent:2020qev,Bollig:2020xdr,Carenza:2020zil,Ferreira:2022xlw,Guo:2023hyp, Fiorillo:2025sln} have set strong constraints in the keV-MeV mass range, they still leave an unexplored region in the parameter space — known as the cosmological triangle for ALP masses between $300~\rm keV$ and $900 ~\rm keV$.
Moreover, cosmological and astrophysical constraints are often highly model-dependent and come with significant uncertainties~\cite{Depta:2020zbh,NEON:2024kwv}, which makes direct terrestrial searches particularly important. 
However, only a few terrestrial experiments probe this range, including Belle II~\cite{Belle-II:2010dht,Dolan:2017osp,Acanfora:2024daq}, Babar~\cite{Dolan:2017osp}, NOMAD~\cite{NOMAD:2000usb}, Beam Dump~\cite{Riordan:1987aw,Mimasu:2014nea,Dobrich:2015jyk,Bauer:2017ris,Harland-Lang:2019zur,Gavela:2019cmq,CCM:2021jmk,Capozzi:2023ffu} and reactor-based experiments~\cite{Dent:2019ueq,AristizabalSierra:2020rom,TEXONO:2006spf,Arias-Aragon:2023ehh,NEON:2024kwv,Dai:2025kai,AristizabalSierra:2025myf}.
 
Nuclear reactors provide a critical experimental platform for studying neutrinos~\cite{FernandezMoroni:2014qlq,SBC:2021yal}, dark matter~\cite{Alfonso-Pita:2021ryw,CONNIE:2024off}, and other new physics phenomena~\cite{Arias-Aragon:2023ehh,TEXONO:2018nir,AristizabalSierra:2020rom,Mirzakhani:2025bqz}, due to their intense particle fluxes and favorable energy conditions. Specifically, reactors can efficiently produce ALPs through the Primakoff process, offering a source of sub-MeV relativistic ALPs.
Experiments employing low-threshold solid-state detectors (e.g., Ge, Si, NaI)~ \cite{Dent:2019ueq,AristizabalSierra:2020rom,TEXONO:2006spf,NEON:2024kwv,Mirzakhani:2025bqz,Dai:2025kai} have been shown to be particularly sensitive to the reactor-produced ALPs.
A promising detection channel in semiconductor detectors is plasmon excitation, a collective mode of electronic excitation~\cite{Gelmini:2020xir,Knapen:2021run,Hochberg:2021pkt} at the eV-scale energies. This channel has been investigated via scattering off target electrons in the context of relativistic dark matter~\cite{Flambaum:2020xxo, Liang:2024xcx,Guo:2024sqh,Sun:2025gyj,Hu:2025dsv} and millicharged particles \cite{CONNIE:2024off,Gong:2025xsd}.

In this work, we propose the detection of reactor-produced ALPs using silicon Skipper-CCDs. We find that the absorption of a keV-scale relativistic ALP by target electrons can efficiently excite the plasmon resonance while emitting an additional MeV-scale energetic photon, in contrast to the dark matter scattering. However, unlike the photon produced via ALP absorption in target nucleus, which can be detected in Ge and NaI detectors~\cite{TEXONO:2006spf,NEON:2024kwv,Dai:2025kai}, this energetic photon escapes from the silicon without depositing energy due to the thinness of the Skipper-CCD sensors. Consequently, the plasmon resonance channel provides an effective and complementary probe for detecting ALPs in the eV-scale energy deposition.
Based on data from the Connie and Atucha-II experiments~\cite{CONNIE:2024off} by the Skipper-CCD sensors situated 30 m from the $3.95~\rm{GW_{th}}$ Angra 2 nuclear reactor and 12 m from the $2.175~\rm{GW_{th}}$ Atucha-II nuclear power plant, respectively, with total exposures of 18.4~g·days and 82~g·days
we present the 90\% confidence level (C.L.) upper limits on the ALP-photon coupling in the  mass range of $0.1$-$ 100$ keV. 
Although the constraints from Connie and Atucha-II are weaker than those from NEON due to their much lower exposures. 
we project that a Connie-like experiment with exposure comparable to NEON could exceed the sensitivity of NEON by a factor of about 1.5, owing to plasmon enhancement in Skipper-CCDs. This demonstrates that plasmon resonances can significantly enhance sensitivity to light ALPs.  Furthermore, we project that  the vIOLETA experiment, with an exposure of 30 kg$\cdot$yr, can surpass the NEON constraint by a factor of three in the same mass range.

\section{Production of ALPs}
Photons are abundantly produced in nuclear reactors. The primary mechanisms responsible for $\gamma$-ray emission include fission, decay of fission products, capture reactions in the fuel and structural materials, inelastic scattering in the fuel, and decay of activation products~\cite{ROOS195998}.
The photon flux for energies above 0.2 MeV can be written as~\cite{bechteler_faissner_yogeshwar_seyfarth_1984},
\begin{equation}
  \frac{d\Phi_{\gamma}}{dE_{\gamma}}=\frac{5.8\times 10^{17}}{\mathrm{MeV\cdot sec}}\left(\frac{P}{\mathrm{MW}}\right)e^{-1.1E_{\gamma}/\mathrm{MeV}}\, , \label{eq:photon_flux}
\end{equation}
where $E_\gamma$ is the photon energy and $P$ is the reactor thermal power in units of $\rm MW$.

Once produced, these photons can interact with the fuel material to produce ALPs. In the mass range of interest, the dominant production mechanism is the Primakoff process on nuclei. The relevant interaction between the ALP field $a$ and the electromagnetic field $A_\mu$ is given by the Lagrangian,
\begin{equation}
\mathcal{L}_{\mathrm{int}} \supset -\frac{1}{4}g_{a\gamma\gamma}aF_{\mu\nu}\tilde{F}^{\mu\nu} \, ,
\end{equation}
where $g_{a\gamma\gamma}$ is the ALP-photon coupling constant, $F_{\mu\nu}$ and $\tilde{F}_{\mu\nu}$ are the electromagnetic field strength tensor and its dual, respectively.

The ALP flux at the detector is obtained by convolving the incident photon flux with the normalized differential cross section,
\begin{equation}\label{eq:flux}
  \frac{d\Phi_{a}}{dE_{a}}=\frac{P_{\mathrm{sur}}}{4\pi L^2}\int _{E_{\gamma}^\mathrm{min}}^{E_{\gamma}^\mathrm{max}}\frac{1}{\sigma_{\mathrm{Tot}}}\frac{d\sigma_{P}}{dE_{a}}(E_{\gamma},E_a)\frac{d\Phi_{\gamma}}{dE_{\gamma}}dE_{\gamma} \, ,
\end{equation} 
where $E^{\mathrm{min}}_{\gamma}$ and $E^{\mathrm{max}}_{\gamma}$ represent the kinetic boundaries, and the total cross section is $\sigma_{\mathrm{Tot}} = \sigma_{\mathrm{SM}} + \sigma_P$. The standard model cross section $\sigma_{\mathrm{SM}}$ for total photon scattering off the core material can be directly obtained from the database~\cite{XCOM}. 
$d\sigma_{\rm P}/dE_a$ denotes the differential Primakoff production cross section~\cite{AristizabalSierra:2020rom,Dai:2025kai}. The survival probability $P_{\mathrm{sur}}$ of the ALP during propagation writes as $P_{\mathrm{sur}} = \mathrm{exp}({-L m_a \Gamma_{2\gamma}/ |{\bf p}_a| })$, where the decay width to two photons is $\Gamma_{2\gamma} = g_{a\gamma\gamma}^2 m_a^3/64\pi$. The distance $L$ is from the reactor to the detector. Finally, we take $E_{\gamma}^{\mathrm{min}}$ = 0.2 MeV and $E_{\gamma}^{\mathrm{max}}$ = 10 MeV, 
as determined by the natural photon threshold in reactor environments. Since the incident photon energies are much smaller than the mass of the target nuclei, the production process is dominant in forward direction and the ALP energy distribution is almost monochromatic $E_a \approx E_\gamma$. 

\begin{figure}[htbp]
    \centering
   \includegraphics[width=\columnwidth]{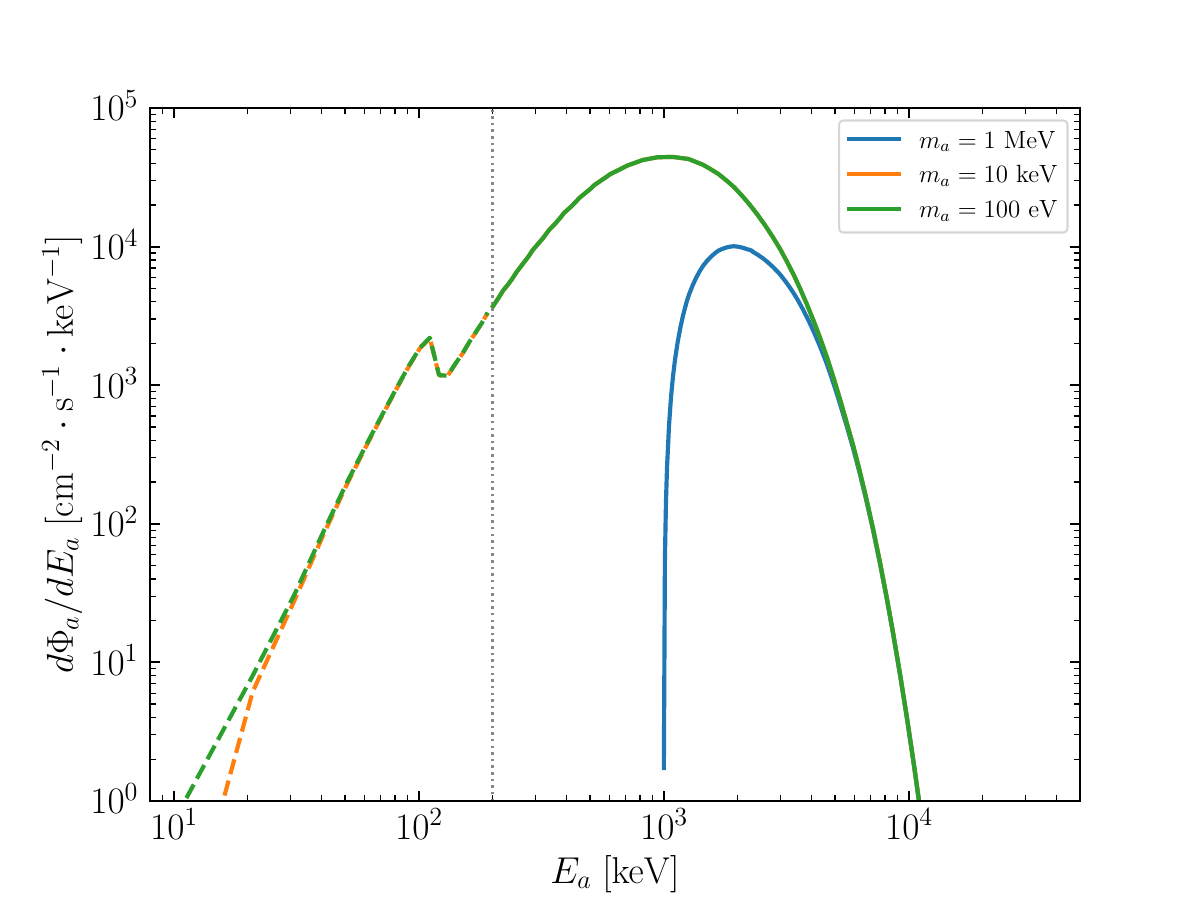}
   \caption{The ALP flux at a detector located 30 m from the $3.95~\rm{GW_{th}}$ nuclear reactor is shown as a function of the ALP energy $E_a$ for ALP masses $m_a =100 ~\rm eV$ (green solid lines), $m_a =10~\rm keV$ (orange solid lines) and $m_a =1 ~\rm MeV$ (blue solid lines). The ALP-photon coupling is set to  $g_{a\gamma\gamma}=10^{-4} ~\rm GeV^{-1}$. 
   The dashed green and orange lines below the 0.2 MeV threshold (vertical dotted line) are obtained by extrapolating Eq.~(\ref{eq:photon_flux}) to lower energies.}
    \label{flux}
\end{figure}

In Fig.~\ref{flux}, 
we show the ALP flux at a detector located 30 m from the $3.95~\rm{GW_{th}}$ nuclear reactor for ALP masses $m_a =100 ~\rm eV$ (green lines), $m_a =10~\rm keV$ (orange lines) and $m_a =1 ~\rm MeV$ (blue lines). 
The vertical dotted line indicates the nominal photon-flux threshold at 0.2 MeV. The dashed green and orange lines below this threshold are obtained by extrapolating the $E_\gamma$ in Eq. \ref{eq:photon_flux} to lower energies. As expected, the reactor-produced ALP flux peaks at MeV-scale energies, so the ALPs remain relativistic when they reach the detector.  We note that the flux in Eq. \ref{eq:flux} receives several kinds of uncertainties, including the escape probability of the photons from the fuel rods ~\cite{ROOS195998}, secondary photon flux from transport and energy loss of the prompt photons in the nuclear core~\cite{CONNIE:2024off}, and possible ALP attenuation or production inside the shielding~\cite{Dent:2019ueq}. Consistent with the treatment in \cite{CONNIE:2024off}, we expect a conservative systematic uncertainty of 10\% in the ALP flux. 

\begin{figure}[htbp]
    \centering
   \includegraphics[width=\columnwidth]{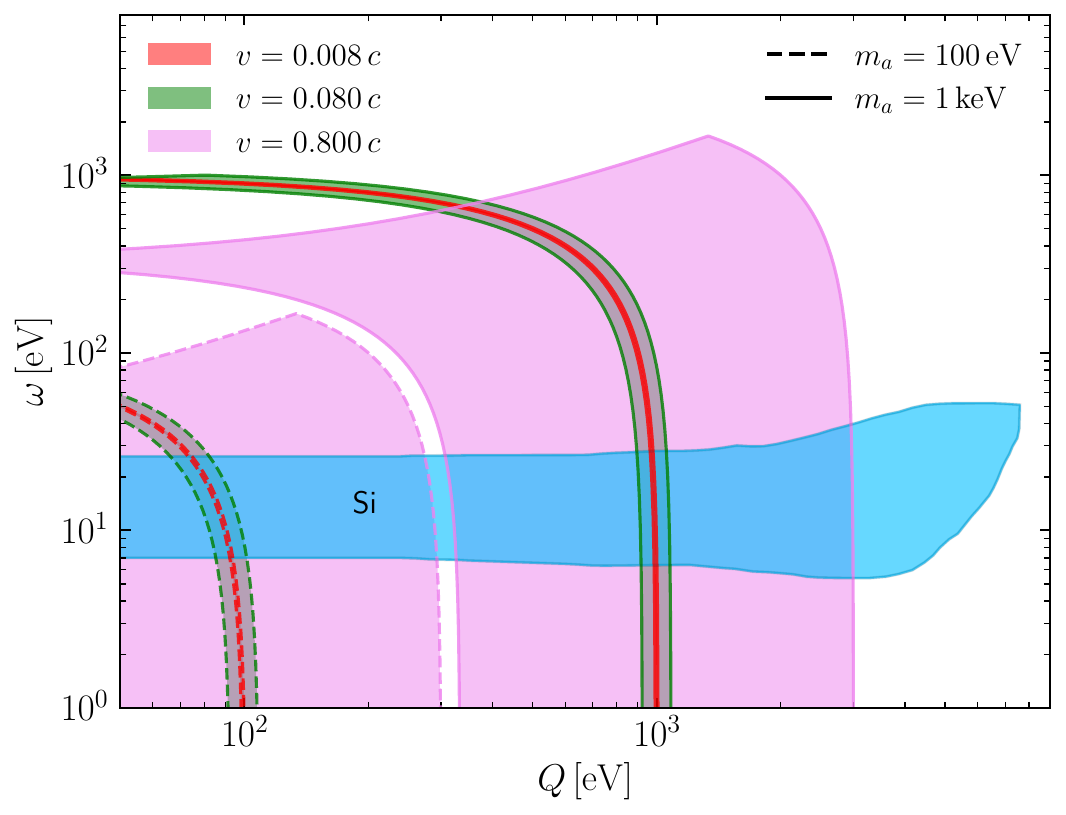}
   \caption{
   The blue region illustrates the plasmonic energy loss function $\rm{Im}[-\epsilon^{-1}(Q, \omega)]>10^{-1}$ for silicon semiconductors on the plane of momentum transfer $Q$ and energy transfer $\omega$. The kinematically accessible regions are shown for 100 eV and 1 keV ALPs with representative incident velocities $v_a=0.008c$ (red), $v_a=0.080c$ (green), and $v_a=0.800c$ (violet). }
    \label{elf}
\end{figure}

\section{Detection of ALPs}
Once relativistic ALPs produced in the reactor reach the detector, they can be detected via the inverse Primakoff process ($a + N \to \gamma + N$) and axion decay, both of which produce energetic photons that serve as appropriate experiment signals, e.g., in Ge and NaI detectors~\cite{Dent:2019ueq,AristizabalSierra:2020rom,Dai:2025kai}. However, as mentioned before, 
these channels are largely inaccessible to silicon Skipper-CCDs, as MeV-scale photons escape the thin sensors without depositing energy. For photons above 40~keV, the attenuation length exceeds the thickness of Skipper-CCDs by more than an order of magnitude, leading to an escape probability well above 90\%. A detailed calculation is given in Appendix B.

We therefore consider the inverse Primakoff-like absorption of the ALP by the target electrons through the exchange of the photon. Owing to their relativistic nature, reactor-produced ALPs can efficiently excite plasmon resonances at the eV-scale energy while emitting an additional MeV-scale photon that escapes the sensors (even for electron recoil instead of collective excitation, the minimum final-state photon energy reaches 0.112~MeV for an incident energy $E_a \approx 0.2$~MeV). In Fig.~\ref{elf},  
we present both the kinematic range of the plasmon resonance in silicon, corresponding to an energy loss function $\mathrm{Im}[-1/\epsilon(Q, \omega)]>10^{-1}$ and the kinematic relation $E_a = \sqrt{(p_a \pm Q)^2} + \omega$ for 100 eV and 1 keV ALPs with given incident velocities ($v_a = 0.008c, 0.08c, 0.8c$).
A pronounced plasmon resonance is observed at energy transfers around $\omega\sim 20 ~\rm eV$ and momentum transfers $Q < 6~\rm keV$. This indicates that relativistic ALPs with masses below 100 keV are capable of exciting plasmons in silicon. We note that for masses above 100 keV, the momentum transfer gradually moves outside the plasmonic region, and the enhancement correspondingly weakens.

Since plasmon excitation requires energies on the scale of $\mathcal{O}$(10 eV), the electrons in the semiconductor remain non-relativistic. Consequently, the electron-photon coupling can be described within the framework of non-relativistic effective field theory \cite{Mitridate:2021ctr},
\begin{eqnarray}\label{leff}
\mathcal{L}_{\mathrm{int}} \supset -eA_0\psi^*_e\psi_e- \frac{ie}{2m_e}{\bf{A}}\cdot(\psi^*_e\overrightarrow{\nabla}\psi_e-\psi^*_e\overleftarrow{\nabla}\psi_e)
+\cdots\,,
\end{eqnarray}
where $\psi_e$ denotes the non-relativistic electron wavefunction.  Eq.~\ref{leff} shows that the second term (the current interaction) is suppressed by the electron mass $m_e$ from the bound state wavefunction.
Therefore, we consider only the contribution from the first term in our analysis.

The ALP-induced electronic transition rate in a semiconductor can be calculated by the energy loss function formalism~\cite{Hochberg:2021pkt,Liang:2024xcx},
which inherently incorporates many-body in-medium effects obtained from first-principles density functional theory (DFT) and has previously been  applied to dark matter-electron scattering~\cite{SuperCDMS:2020ymb}. Here, we extend this framework to reactor ALPs. For an ALP characterized by a mass $m_a$ and momentum $\mathbf{p}_a$, the transition rate per unit volume is given by
\begin{eqnarray}\label{eq:transition rate}
\Gamma({\bf p}_a)=\int\frac{d^3{\bf Q}}{(2\pi)^3}|V({\bf Q},\omega)|^2\left[2\frac{Q^2}{e^2}{\rm Im}\left(-\frac{1}{\epsilon({\bf Q},\omega)}\right)\right]\,,
\end{eqnarray}
where $e$ is the electron charge and ${\bf Q}$ denotes the three-momentum transfer to the electron. The transfer energy is defined by $\omega=E_a-E_\gamma$, where $E_a$ and ${\bf p}_a$ are the energy and momentum of the incoming ALP, and $E_\gamma$ is the energy of the outgoing photon that escapes from the sensors.
For the isotropic target materials, the energy loss function can be simplified to $\rm{Im}[-\epsilon^{-1}(\bf{Q}, \omega)]=\rm{Im}[-\epsilon^{-1}(Q, \omega)]$, and thus the transition rate becomes $\Gamma({\bf p}_a)=\Gamma(p_a)$. While the crystal anisotropy effects could provide daily modulation signatures as a useful signal discriminant \cite{Boyd:2022tcn} and are referred to further work.
We use the DarkELF package (GPAW model) to compute the energy loss function for energy transfers below 50 eV and momentum transfers below the Fermi momentum $k_F \sim 5$ keV, as this approach offers a more accurate description in this regime (see Refs.~\cite{Knapen:2021bwg,Essig:2024ebk}) and agrees with electron energy loss spectroscopy (EELS) data. The target material response is
dominated by low momentum transfers $k <k_F $ and the energy loss function is less appropriate in higher momentum regime. In Eq.~\ref{eq:transition rate}, the potential $V(\mathbf{Q},\omega)$ for the ALP-photon interaction via the $g_{a\gamma\gamma}$ coupling is given by 
\begin{align}
|V({\bf Q})|^2&=\left(\frac{g_{a \gamma\gamma} e}{q^2}\right)^2[E_a(E_a-\omega)(m_a^2+q^2)\\
&-m_a^2(E_a-\omega)^2-(m_a^2+q^2)^2/4]/[4E_a(E_a-\omega)]\,,\nonumber
\end{align}
where $q^2=Q^2-\omega^2$. The differential event rate per unit volume is given by the convolution of the ALP flux (Eq.~\ref{eq:flux}) with the transition rate (Eq.~\ref{eq:transition rate})

\begin{align}
\frac{dR}{d\omega}=\int\frac{dE_a}{\rho_T} \int\frac{d\Omega}{4\pi}\frac{d\Phi_a}{dE_a}\frac{E_a}{p_a}\Gamma(p_a)\delta(\omega+E_\gamma-E_a)\,,
\end{align}
where  $\rho_T$ is the mass density of the semi-conductor target. 
A detailed calculation of the event rate can be found in the Appendix A.  

\section{Experimental sensitivity}
The silicon Skipper-CCD detector represents a significant advance in low-threshold particle detection, providing non-destructive readout with single-electron sensitivity and excellent background control~\cite{Tiffenberg:2017aac}. 
The Coherent Neutrino-Nucleus Interaction Experiment (Connie) has pioneered its application at the nuclear reactor~\cite{CONNIE:2019swq}. 16 Skipper-CCD sensors are located outside the dome of the $3.95~\rm{GW_{th}}$ Angra 2 nuclear reactor in Brazil, approximately 30 m from the core \cite{CONNIE:2024pwt}. Two of the Skipper-CCD, each with an active mass of 0.247 g, reach an energy threshold as low as 15 eV,
corresponding
to about four electrons on average. Below 15 eV, the signal would be affected by additional on-chip noise sources (e.g., dark current, spurious charge, and clock-induced charge backgrounds) that could produce fake events~\cite{CONNIE:2024pwt} in 1–2 electron events. With an exposure of 18.4 g$\cdot$days from 2021 to 2023, the experiment recorded a reactor-OFF background rate of approximately $4000~\rm{kg^{-1}~day^{-1}~keV^{-1}}$ \cite{CONNIE:2024off}. For the Atucha-II experiment, a 2.2 g Skipper-CCD is located 12 m from a $2.175~\rm{GW_{th}}$ pressurized heavy water reactor operating with natural $\rm{UO_2}$ fuel in Buenos Aires Province, Argentina. During the 2023 operational run, it recorded the data with 82 g-days of exposure and the measured background rate under reactor-OFF conditions was approximately $30000~\rm{kg^{-1}~day^{-1}~keV^{-1}}$ \cite{CONNIE:2024off}. 

The Neutrino Interaction Observation with a Low Energy Threshold Array experiment (vIOLETA) is designed to search for neutrino interactions using a large array of novel silicon CCDs~\cite{Fernandez-Moroni:2020yyl,Fernandez-Moroni:2021nap}. These devices feature a low detection threshold of 15 eV and are located 12 m from the core of a nuclear reactor, operating at a steady-state thermal power of 2~$\rm{GW_{th}}$, consistent with the Atucha II Nuclear Power Plant. It has a fiducial mass of 10 kg and a data-taking period of 3 years~\cite{Fernandez-Moroni:2020yyl}. Assuming 45 days of reactor-off data taking per year, the reactor-on exposure is 26.3 kg$\cdot$yr. In this work, we adopt a conservative background rate of $1000~\rm{kg^{-1}~day^{-1}~keV^{-1}}$, 
 although future upgrades may reduce it to $100~\rm{kg^{-1}~day^{-1}~keV^{-1}}$~\cite{Fernandez-Moroni:2020yyl}.
Notably, in large-scale CCD implementations, the above-mentioned high-energy final-state photons may not escape the entire detector volume and can produce the 
 secondary $\gamma$ and electron products by interacting within the detector or surrounding materials. The energy deposition of these energetic secondary products will produce extended tracks or point-like events in the CCD, accompanied by the single- and few-electron events through the Cherenkov and recombination photoabsorption~\cite{Du:2020ldo,Du:2023soy}. These correlated low-energy events can be vetoed by the halo mask~\cite{Du:2020ldo}. Moreover, high-energy $\gamma$ interactions are expected to produce energetic ionization tracks, which can be vetoed using coincidence and anti-coincidence in multilayer CCD detectors. A quantitative estimate of both the veto efficiencies and possible residual low energy events is detector-design dependent and requires a full detector-geometry simulation with GEANT4 or FLUKA.

\begin{figure}[h]
    \centering
     \includegraphics[width=\columnwidth]{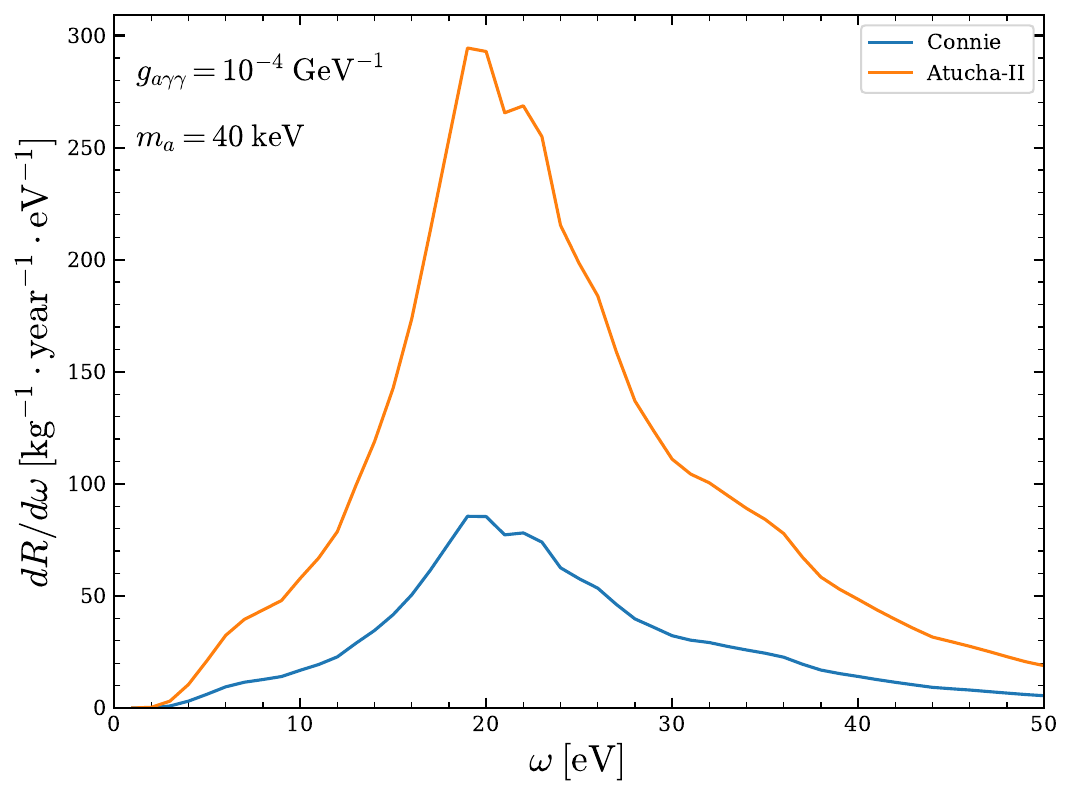}
    \caption{The differential event rate $dR/d\omega$ as a function of the deposited energy $\omega$ for an ALP with mass $m_a=40~\rm{ keV}$ and a coupling constant of $g_{a\gamma\gamma}=10^{-4}~\rm{GeV}^{-1}$, produced via the Primakoff process. The blue and orange lines correspond to CONNIE and Atucha-II, respectively. }
    \label{rate_gae}
\end{figure} 

The differential event rates for a 40 keV ALP, with a coupling constant of $g_{a\gamma\gamma}=10^{-4}~\mathrm{GeV}^{-1}$, are shown in Fig.~\ref{rate_gae}. The blue and orange lines correspond to the Connie, and Atucha-II experiments, respectively.
As expected, the plasmon resonance appears in the low-energy region for light ALPs, exhibiting a peak in the differential event rate around 20 eV. The difference in the observed differential event rate is attributable to the different ALP fluxes from the reactors at each experiment. The Atucha-II experiment, with its higher flux, predicts a larger differential rate than the Connie experiment.

In this work, we present 90\% C.L. exclusion limits on the ALP-photon coupling $g_{a\gamma\gamma}$ from an analysis of data from reactor  experiments. To obtain the expected signal event spectrum, the theoretical differential event rate was convolved with the detection efficiency as a function of energy, as provided in Refs.~\cite{CONNIE:2024pwt,Depaoli:2024bgs}, thereby correcting for detector effects. Following this correction, the sensitivity was derived using a likelihood test based on a single energy bin. For the Connie experiment, which covers an energy range from 15 eV to 215 eV, this yields a upper limit of 6.2 signal events with 90\% C.L.~\cite{CONNIE:2024off}. A similar analysis for the Atucha-II experiment, performed over a single bin spanning 40 eV to 240 eV, sets a upper limit of 30.9 signal events~\cite{CONNIE:2024off}. 

\begin{figure}
    \centering
    \includegraphics[width=\columnwidth]{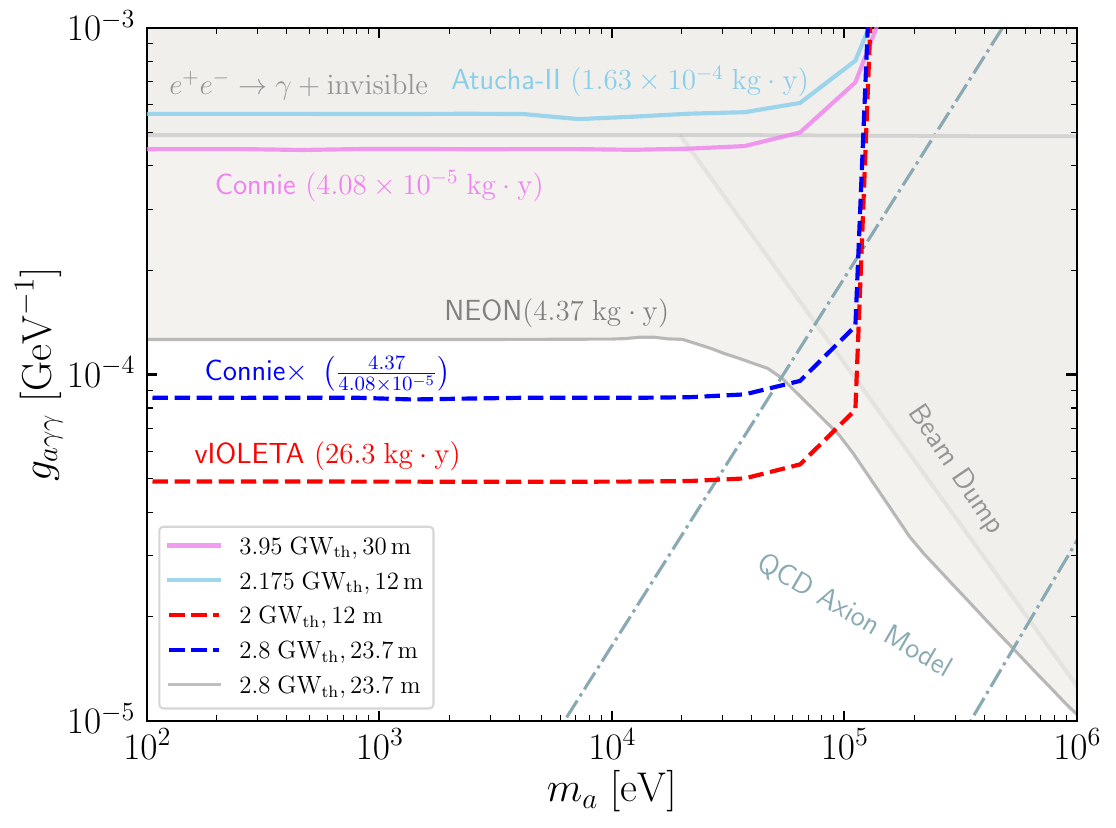}
    \caption{
    The 90\% C.L. upper limits on
    the ALP-photon coupling as a function of the ALP mass. The blue solid, purple solid, and red dashed lines represent the limits from the Connie, Atucha-II, and vIOLETA, respectively. For comparison, we also include existing limits from beam dump experiments~\cite{Bechis:1979kp,1987Search,Bjorken:1988as,Blumlein:1991xh,Andreas:2010ms,CCM:2021jmk}, the NEON experiment~\cite{NEON:2024kwv}, $e^{+}e^{-}\rightarrow \gamma+\rm{invisible}$~\cite{Jaeckel:2015jla} and the parameter space favored by a representative QCD axion model~\cite{DiLuzio:2020wdo}.}
    \label{fig:example}
\end{figure}

In Fig.~\ref{fig:example}, we show the constraints from the Connie and Atucha-II experiments. For the sake of
comparison, selected constraints from
other terrestrial detection experiments, including beam dump experiments~\cite{Bechis:1979kp,1987Search,Bjorken:1988as,Blumlein:1991xh,Andreas:2010ms,CCM:2021jmk}, the NEON experiment~\cite{NEON:2024kwv}, $e^{+}e^{-}\rightarrow \gamma+\rm{invisible}$~\cite{Jaeckel:2015jla} and the parameter space favored by a representative QCD axion model~\cite{DiLuzio:2020wdo} are also shown.
As shown in Fig.~\ref{fig:example}, 
for ALP masses below 100 keV, the constraints from Connie and Atucha-II are weaker than the best existing limit from NEON experiment. This is primarily because these reactor-based measurements have significantly smaller exposures. Above 100 keV, the sensitivities are suppressed since the momentum transfer gradually exceeds the range allowed by the plasmon excitation in the detector material, which suppresses the plasmon resonance. Notably, although the differential event rate of Atucha-II is higher than that of Connie, its final constraint is weaker because the absence of significant plasmon enhancement in the relevant energy range. 
To illustrate the enhancement from plasmon excitation, we also present in the same figure the projected sensitivity of a Connie-like detector
with exposure increased to 1596 kg$\cdot$days during reactor-on period as the NEON experiment. For a equivalent comparsion, we also adopt the reactor flux and baseline, i.e., $P = 2.8$~GW$_{\rm th}$ and $L=23.7$~m corresponding to the NEON setup to ensure the same flux at the detector. The background rates and the systematic uncertainties are conservatively assumed to equal the level of the present Connie experiment, since increasing the exposure and upgrading the detectors and shielding in the future are expected to decrease both of them. At the comparable flux and exposure, the use of Skipper-CCDs with plasmon excitation improves the sensitivity by a factor of 1.5 relative to the limit derived from NEON, which uses a NaI solid-state detector. Aside from the aforementioned systematic uncertainty from the axion flux, there are contributions from the energy loss function $<20\%$~\cite{Essig:2024ebk}, energy resolution $<8\%$ and the efficiency curve of the detector $2\%$. As a result, the overall systematic uncertainty on the exclusion limit is 9.5\%. See \cite{CONNIE:2024off} for a detailed discussion on these uncertainties.
Assuming an energy bin spanning 15–215 eV and a conservatively estimated detection efficiency (adopting the same value as Connie), the projected exclusion limit from vIOLETA experiment exceeds the current best constraints from NEON by a factor of three, as shown in Fig.~\ref{fig:example}. This improvement demonstrates the potential of combining reactor sources with the plasmon-excitation detection channel.



\section{Conclusion and outlook}
In this work, we investigate the production of ALPs in nuclear reactors through the Primakoff process with target nuclei. Subsequently, ALPs entering the detector can induce collective electron excitation in silicon. Based on low-threshold data from the Connie and Atucha-II experiments, we set constraints on the ALP–photon coupling $g_{a\gamma\gamma}$. Our results demonstrate the viability of plasmon-based detection as a channel for probing in the mass range of 0.1-100 keV ALPs. We project that the vIOLETA experiment, with a total exposure of 30 kg$\cdot$yr, could enhance the sensitivity to the ALP–photon coupling by a factor of three, surpassing the current limits set by the NEON experiment. By leveraging the plasmon excitation mechanism, vIOLETA experiment offers a versatile platform
for ALP searches. Furthermore, the methodology developed in this work can be systematically extended to other ALP interaction channels, opening new pathways for terrestrial experiments to previously inaccessible regions of the ALP parameter space.

\section*{Acknowledgments}
This work is supported by the National Natural Science Foundation of China (NNSFC)  No. 12275134, NNSFC No. 12305111, and Jiangxi Provincial Natural Science Foundation 20252BAC200168. L. Su is supported by the Alexander von Humboldt Foundation.


\appendix

\section{Final-State Photon Escape Probability}
\label{App:A}
The scattering rate $\Gamma(p_a)$ of an electron from the ground state $|0\rangle$ in semi-conductor target induced by an ALP with the momentum $p_a$ can be obtained by the Fermi’s golden rule or equivalently from the Born approximation in the context of scattering theory
\begin{align}
\Gamma(p_a)=\sum_f\int\frac{d^3p_\gamma}{(2\pi)^3}\frac{|M_{i\rightarrow f}|^2}{4E_a E_\gamma}2\pi\delta(\omega_f+E_\gamma-E_a)\,,
\end{align}
where $\omega_f$ is the energy of final state $|f\rangle$ of electron under the assumption of zero energy ground state. Meanwhile,
the transfer energy is defined by the $\omega=E_a-E_\gamma=\sqrt{p_a^2+m_a^2}-\sqrt{|{\bf p}_a-{\bf Q}|^2}$ with $p_a =|p_a|$ and transfer
momentum Q. ALP interacts with electrons via a photon mediator, and the amplitude $M_{i\rightarrow f}$ is related to the T-matrix via
\begin{eqnarray}
\left<{\bf p}_\gamma,f | iT|{\bf p}_a,0\right>&=&iM_{i\rightarrow f}2\pi\delta(\omega_f+E_\gamma-E_a)\nonumber\\
&=&\epsilon^*_{\mu}\varepsilon^{\mu 0\alpha\beta}p_{3\alpha}q_{\beta}\frac{ig_{a \gamma\gamma} e}{q^2}\nonumber\\
&&\left<f|\hat{\rho}(\bf Q)|0\right>2\pi\delta(\omega_f-\omega)\,,
\end{eqnarray}
where $\hat{\rho}({\bf Q})=\int d^3 {\bf x} \sum_i\delta({\bf x}-\hat{r}_i) e^{-i{\bf Q} \cdot{\bf x}}=\sum_i e^{-i {\bf Q}\cdot\hat{\bf r}_i}$
is the momentum-space electron density operator, with $\hat{\bf r}_i$ being the position operators of relevant electrons in the target. The energy loss function can be expressed in terms of
the electron density operator as follows
\begin{eqnarray}
{\rm Im}\left(-\frac{1}{\epsilon({\bf Q},\omega)}\right)=\frac{\pi e^2}{Q^2}\sum_f|\left<f|\hat{\rho}({\bf Q})|0\right>|^2\delta(\omega_f-\omega)\,,
\end{eqnarray}
where $e^2=4\pi\alpha$. Therefore, the scattering rate can be rewritten as
\begin{eqnarray}\label{rate}
\Gamma(p_a)&=&\sum_f\int\frac{d^3p_\gamma}{(2\pi)^3}\frac{|M_{i\rightarrow f}|^2}{4E_a E_\gamma}2\pi\delta(\omega_f+E_\gamma-E_a)\nonumber\\
&=&\int\frac{d^3{\bf Q}}{(2\pi)^3}\left(\sum_{s,s'}\frac{|\epsilon_{\mu}\varepsilon^{\mu 0\alpha\beta}p_{3\alpha}q_{\beta}|^2}{4E_a E_\gamma}\right)\left(\frac{ig_{a \gamma} e}{Q^2-\omega^2}\right)^2\nonumber\\
&&\sum_f|\left<f|\hat{\rho}(\bf Q)|0\right>|^22\pi\delta(\omega_f-\omega)\nonumber\\
&=&\int\frac{d^3{\bf Q}}{(2\pi)^3}|V({\bf Q})|^2\left[2\frac{Q^2}{e^2}{\rm Im}\left(-\frac{1}{\epsilon({\bf Q},\omega)}\right)\right]\,,
\end{eqnarray}
where $E_\gamma=E_a-\omega$, the potential $V({\bf Q},\omega)$ for the interaction of the ALP with the electron via a photon mediator in momentum space is defined as
\begin{align}
|V({\bf Q}, \omega)|^2&=\left(\frac{g_{a \gamma\gamma} e}{q^2}\right)^2[E_a(E_a-\omega)(m_a^2+q^2)-m_a^2\\
&(E_a-\omega)^2
-(m_a^2+q^2)^2/4]/[4E_a(E_a-\omega)]\,.\nonumber
\end{align}
The event rate for collective excitations induced by an ALP flux is given by Eq. (\ref{rate}). Given the isotropy of the target material (i.e., the independence of ${\rm Im}[-1/\epsilon({\bf Q},\omega)]$ on the direction of $\bf Q$), we evaluate the integral by aligning the initial ALP momentum $\bf p_a$ with the z-axis of a spherical coordinate system and integrating over the polar angle of $\bf Q$, leading to:
\begin{eqnarray}
\frac{dR}{d\omega}&=&\frac{1}{\rho_T}\int dT_a\int\frac{d\Omega}{4\pi}\frac{d\Phi_a}{dT_a}(\frac{E_a}{p_a})\Gamma(p_a)\delta(\omega+E_\gamma-E_a)\nonumber\\
&=&\frac{1}{\rho_T}\int\frac{Q^2dQd\cos\theta d\phi}{(2\pi)^3}\int dE_a\frac{d\Phi_a}{dE_a}\frac{E_a}{p_a}|V({\bf Q}, \omega)|^2\nonumber\\
&&\left[2\frac{Q^2}{e^2}{\rm Im}\left(-\frac{1}{\epsilon({Q},\omega)}\right)\right]\delta(\omega+E_\gamma-E_a)\nonumber\\
&=&\frac{1}{\rho_T}\frac{2}{e^2}\int\frac{Q^2dQ}{(2\pi)^2}\int dE_a\frac{d\Phi_a}{dE_a}\int\left|\frac{d\cos\theta}{dE_\gamma}\right|dE_\gamma\frac{Q^2E_a}{p_a}\nonumber\\
&&\left(\frac{g_{a \gamma} e}{Q^2-\omega^2}\right)^2\bigl[E_a(E_a-\omega)(m_a^2+Q^2-\omega^2)-m_a^2\nonumber\\
&&(E_a-\omega)^2-(m_a^2+Q^2-\omega^2)^2/4\bigl]\left[4E_a(E_a-\omega)\right]\nonumber\\
&&\times~ {\rm Im}\left(-\frac{1}{\epsilon({Q},\omega)}\right)\delta(\omega+E_\gamma-E_a)\nonumber\\
&=&\frac{1}{\rho_T}\frac{1}{2\pi^2}\int Q^3dQ\int dE_a\frac{d\Phi_a}{dE_a}\left(\frac{g_{a \gamma}}{Q^2-\omega^2}\right)^2\frac{1}{4p_a^2}\nonumber\\
&&\bigl[E_a(E_a-\omega)(m_a^2+Q^2-\omega^2)-m_a^2(E_a-\omega)^2-\nonumber\\
&&(m_a^2+Q^2-\omega^2)^2/4\bigl]{\rm Im}\left(-\frac{1}{\epsilon({Q},\omega)}\right)\Theta[E_\gamma-E^\pm]\,,\nonumber\\
\end{eqnarray}
where $\rho_T$ is the mass density of semiconductor target and $E^\pm (p_a) =  \sqrt{(p_a\pm Q)^2}$. In the first line, we insert  an identity $1 =\int d\omega \delta(E_\gamma-E_a+\omega)$ to introduce the variable of deposited energy $\omega$ and the $dT_a = d(E_a-m_a) = dE_a$ is considered in second line. Then we take a variable transformation from $\cos\theta_{{\bf Q p}_a}$ to $E_{p_\gamma}$, along with its corresponding Jacobian
\begin{eqnarray}
\left|\frac{d\cos\theta_{{\bf Q p}_a}}{dE_{\gamma}}\right|=\left(\frac{E_{\gamma}}{p_aQ}\right)\,.
\end{eqnarray}
The step function in the last line indicates whether the span of integration over $E_{\gamma}$ covers the zero in the delta function.

\section{Final-State Photon Escape Probability}
\label{App:B}
For a photon of energy $E_\gamma$ in the target material, the attenuation length is given by
\begin{eqnarray}
L=\frac{1}{\mu}=\frac{1}{\rho\cdot(\mu/\rho)}\,,
\end{eqnarray}
where $\mu$ is the linear attenuation coefficient, $\rho$ is the material density, and $\mu/\rho$ is the mass attenuation coefficient,
which is available from the NIST XCOM database~\cite{hubbell2004tables}.
The mass attenuation coefficient is related to the total photon cross section $\sigma_{\rm{tot}}$ per atom by,
\begin{equation}
\mu/\rho = \sigma_{\rm tot} / (u A),
\end{equation}
where $u$ is the atomic mass unit and $A$ is the relative atomic mass of the target element. 
The total cross section can be written as the sum over contributions from the following photon interactions~\cite{hubbell2004tables},
\begin{eqnarray}
\sigma_{\rm{tot}}=\sigma_{\rm{pe}}+\sigma_{\rm{coh}}+\sigma_{\rm{incoh}}+\sigma_{\rm{pair}}+\sigma_{\rm{trip}}\,,
\end{eqnarray}
where $\sigma_{\rm{pe}}$ is the atomic photoelectric cross section, $\sigma_{\rm{coh}}$ and $\sigma_{\rm{incoh}}$ are the coherent (Rayleigh) and incoherent (Compton) scattering cross sections, respectively, and $\sigma_{\rm{pair}}$ and $\sigma_{\rm{trip}}$ are the cross sections for electron-positron production in the fields of the nucleus and of the atomic electrons. The main contributions to the cross section in the relevant energy range of this work are from the photoelectric effect and incoherent Compton scattering.

The photon escape probability through a silicon detector can be written as,
\begin{eqnarray}
P=e^{-\mu d}=e^{ -d/L}\,,
\end{eqnarray}
\begin{figure}
 [htbp]
    \centering
    \includegraphics[width=\columnwidth]{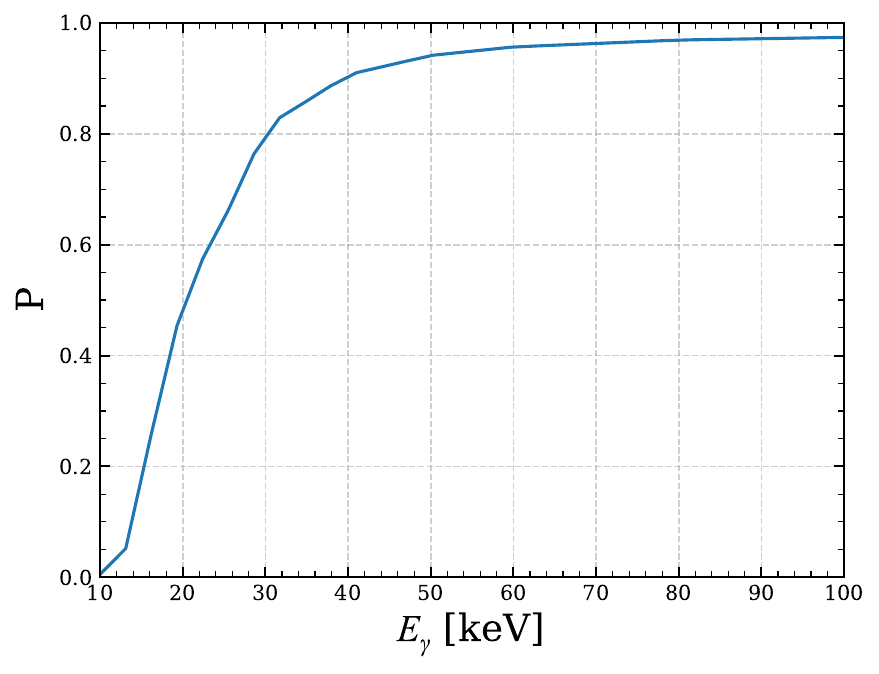}
    \caption{Escape probability for photons propagating through a silicon detector as a function of $E_\gamma$.}
    \label{fig:attenuation}
\end{figure}
where $d$ is the detector thickness. For the Connie, Atucha-II and vIOLETA experiments, which employ Skipper-CCD sensors with thickness $d = 675~\mu\mathrm{m}$. Using the density of silicon $\rho = 2.33~\mathrm{g/cm^3}$, the amplitude of photon attenuation length $L$ is derived from the mass attenuation coefficient $\mu/\rho$, taking from NIST XCOM database~\cite{hubbell2004tables}. 
The resulting escape probability for photons propagating through a silicon detector, as a function of $E_{\gamma}$, is shown in Fig.~\ref{fig:attenuation}.

As shown in Fig.~\ref{fig:attenuation}, the escape probability exceeds 90\% for final-state photon energies above 40 keV. Since the threshold energy of the final-state photons relevant to this work lies well above 40 keV, the detector response to such photons is negligible and not considered further.

\bibliography{refs}

\end{document}